\documentstyle[aps,preprint,epsfig]{revtex} 
\tightenlines
\begin{document}


\draft

\title{New Limits on Doubly Charged Bileptons from LEP Data, and Search at
Future Electron--Positron and Electron--Photon Colliders} 

\author{E.\ M.\ Gregores$^{1,2}$, A.\ Gusso$^{1,3}$ and S.\ F.\ Novaes$^1$}

\address{\vspace{1ex}
$^1$ Instituto de F\'{\i}sica Te\'orica, Universidade  Estadual Paulista, \\  
01405--900 S\~ao Paulo -- SP, Brazil\\[0.5ex]
$^2$ Department of Physics, University of Wisconsin \\
Madison, WI 53706, USA \\[0.5ex]
 $^3$ Instituto de F\'{\i}sica Corpuscular - C.S.I.C., Universitat de Val\`encia, \\
E-46071 Val\`encia, Spain}

\maketitle

\begin{abstract}
We show that the accumulated LEP-II data taken at $\sqrt{s} =$ 130 to 206
GeV can establish more restrictive bounds on doubly charged bilepton
couplings and masses than any other experiment so far.  We also analyze
the discovery potential of a prospective linear collider operating in both
$e^+ e^-$ and $e \gamma$ modes. 
\end{abstract}
 

\pacs{14.80.-j, 12.60.-i, 13.88.+e}

\newpage 

\section{Introduction}
\label{sec:intr}

Bileptons are bosons carrying double leptonic number that are predicted by
some extensions of the Standard Model like $SU(15)$ based GUT theory
\cite{su15}, 3-3-1 model \cite{331}, and left--right symmetric models
\cite{lr}. They can either be scalar or vector particles with different
charges, {\it i.e.} neutral, singly, or doubly charged. In the 3-3-1
model, the symmetry breaking $SU(3)_L \times U(1)_X \to SU(2)_L \times
U(1)$ is responsible for the appearance of vector bileptons, while in the
left-right model, based on $SU(2)_L \times SU(2)_R \times U(1)_{B-L}$, the
bileptons correspond to the doubly charged scalar bosons present in the
symmetry breaking sector.

Bileptons can participate in a large variety of processes, both at low and
high energies. However, no signal has yet been found, and bounds on their
masses and couplings could be obtained from the analysis of lepton number
violating processes \cite{leptviolation}, and muonium--antimuoniuon
($M$--$\overline{M}$) conversion \cite{muonanti,Cuypers} experiments. The
existing mass limits for both singly and doubly charged bosons
\cite{muonanti,modeldep} are all model dependent. These bounds still allow
the existence of low--mass bileptons with a small coupling constant. The
limits from high energy experiments \cite{Cuypers} like $e^+ e^-$
collisions are, in principle, less restrictive than the low energy bounds.

In this paper we explore the production of doubly charged bileptons in
both $e^+ e^-$ and $e \gamma$ colliders, looking for the most promising
signature for identifying these particles: an isolated same sign, planar,
and $p_T$ balanced two muons(antimuons) event
\cite{Cuypers,Cuypers2,Frampton}. We made a model independent analysis,
working in the context of a general effective $SU(2)_L \times U(1)_Y$
Lagrangian that couples bileptons to leptons \cite{Cuypers}
\begin{equation}
{\cal L} = \lambda_1 \bar{\ell}^c i \sigma_2 \ell L_1
+ \tilde{\lambda}_1 \overline{e}^c e \tilde{L}_1 
+ \lambda_2 \overline{\ell}^c \gamma^{\mu} e L_{2 \mu} 
+ \lambda_3 \overline{\ell}^c i \sigma_2 \vec{\sigma} \ell \cdot
\vec{L}_3 + {\rm h.c.} \;,
\label{lagrangian}
\end{equation}
where $\ell = (e_L, \nu_L)$ are left--handed $SU(2)_L$ lepton doublets,
and $e = e_R$ are right--handed charged singlet leptons. The charge
conjugated fields are defined as  $\overline{\ell}^c = (\ell^c)^{\dagger}
\gamma^0 = -\ell^TC^{-1}$.   The subscript of the bilepton fields $L =1,
2, 3$ indicates whether  they are singlet, doublet, or triplet under
$SU(2)_L$. The terms concerning the doubly charged bileptons can be
written as
\begin{equation}
{\mathcal L}^{--} = 
\tilde{\lambda}_1^{ij}\, \overline{e}^c_i P_R\, e_j \tilde{L}_1^{--}  
+ \lambda_2^{ij}\, \overline{e}^c_i \gamma^{\mu} P_R\, e_j L_{2 \mu}^{--} 
+ \sqrt{2}\, \lambda_3^{ij}\, \overline{e}^c_i P_L\, e_j L_3^{--}
+ {\rm h.c.} \; ,
\end{equation}
where $e_i$ represent the charged leptons with flavor indices $i,j = 1, 2,
3$, and $P_{R(L)}=(1\pm\gamma_5)/2$ are the helicity projectors. We will
consider here only flavor diagonal bilepton couplings since very
restrictive bounds are imposed by low--energy experiments when flavor
violation can take place \cite{leptviolation,Cuypers}. These couplings
will also be considered real to avoid CP violating processes.

\section{Limits from LEP} 
\label{sec:lep} 

The whole LEP program has been very successful. They were able to achieve 
both energy and luminosity beyond the values initially expected. However
no evidence for new physics has yet been found.  From the non observation
of    pair produced bileptons model independent lower limits on the masses
of these particles can be established to be  close to 100 GeV. However,
the present most stringent bounds on the mass and coupling of doubly
charged bileptons come from the results of muonium--antimuonium conversion
experiments \cite{muonanti}. For flavor diagonal couplings, these
measurements require that the ratio of the bilepton coupling and its mass
must satisfy $\tilde{\lambda}_1/M_B < 0.20$ TeV$^{-1}$ ($90\%$ C.L.),
$\lambda_2/M_B < 0.27$ TeV$^{-1}$ ($95\%$ C.L.), and $\lambda_3/M_B <
0.14$ TeV$^{-1}$ ($90\%$ C.L.).

We found that these limits can be overridden for the mass range
kinematically accessible at LEP collider. The occurrence of a high--energy
event presenting just a $p_T$ balanced co--planar pair of same--sign
leptons, notably muons or antimuons, would be a striking evidence for the
presence of a bilepton. Since there is no Standard Model background for
this kind of process, the observation of one of such dimuon event would
already constitute a important step towards the bilepton discovery. From
the non observation of such event a $90\;(95)\%$ C.L. upper bound on the
values of $\lambda$ and $M_B$ can be obtained based on the predited
number of $2.3 \;(3.0)$ events.  

This signature results predominantly from the process depicted in Fig.\
\ref{eefig}(a), and its cross-section has been evaluated in the
equivalent particle approximation \cite{EPA}. The deviation from the exact
calculation is expected to be small since this process is dominated by
events where the incident particle is scattered at a very small angle. The
relevant cross-section is given by,
\begin{equation}
 \sigma(E_{e^+},\hat{s})_{e^+ e^- \rightarrow \mu^- \mu^-} 
= \int_{x_{\rm min}}^1 dx\, F^{e^-}_{e^+}(x, E_{e^+})\, 
\sigma(\hat{s})_{e^- e^- \rightarrow \mu^-\mu^-}  ,
\label{xsection}
\end{equation}
where $F^{e^-}_{e^+} (x, E_{e^+})$ is the equivalent electron distribution
function of the initial positron. It gives the probability that an
electron with energy $E_{e^-} = x E_{e^+}$ is emitted from a positron beam 
with energy $E_{e^+}$. The same holds true for the positron contents of the
electron. This distribution is \cite{pythia}
\begin{equation} 
F^{e^{-}}_{e^{+}} (x, E_{e^{+}}) = \frac{1}{2} \left \{
\frac{\alpha}{2 \pi} \left[ \ln \left( \frac{E_{e^{+}}}{m_e}
\right)^2 - 1 \right]  \right \}^2 \left(
\frac{1}{x} \right) \left( \frac43 + x - x^2 - \frac43 x^3 + 2 x
(1+x)\ln x \right). 
\label{eedist}
\end{equation}
The cross-section for the subprocess $e^- e^- \rightarrow \mu^- \mu^-$ 
is
\begin{equation}
\sigma(\hat{s}) = 
S \frac{\lambda^{4}\hat{s}}
{24 \pi[(\hat{s} - M_B^2)^2 + M_B^2 \Gamma^2]} \;,
\end{equation}
where $\Gamma = G\lambda^{2}M_B/(8\pi)$ is the $s$-channel resonance
width  for a bileptons of mass $M_B$, $\hat{s}=xs$ is the subprocess
squared center of mass energy,  $S=3,\,1,\,12$ and $G=3,\,1,\,6$ for 
$\tilde{L}_{1}^{--},\, L_{2 \mu}^{--}$ and $L_{3}^{--}$, respectively.

We present in Tab.\ \ref{tableLEP} the integrated luminosities for the
different LEP energies we used in our calculations \cite{LEPluminosity}.
The total number of expected events (pairs of muons or antimuons with
total invariant mass $M_B$) is calculated considering  the luminosities
obtained at center of mass energies that are larger than the bilepton
mass, {\it i.e.},
\begin{equation}
N(M_B)=2\sum_i \Theta(\sqrt{s_i}-M_B)\,{\mathcal L}(\sqrt{s_i})\,
\int_{\hat{s} = 4 m_\mu^2}^{s_i}d\hat{s} \; \sigma(\sqrt{s_i}/2,\hat{s})_{e^+  
e^- \rightarrow   \mu^- \mu^-} 
\end{equation}
where $N(M_B)$ is the number of expect events with mass $M_B$, ${\mathcal
L}(\sqrt{s_i})$ assumes the values in Tab.\ \ref{tableLEP}.          The
factor $2$ stands for the fact that $\sigma_{e^+ e^- \rightarrow \mu^-
\mu^-} = \sigma_{e^+ e^- \rightarrow \mu^+ \mu^+}$.

In Fig.\ \ref{LEP}(a) and (b) we present, respectively, the limits on
$L_{2\mu}^{--}$ and  $L_{3}^{--}$ coupling constant that are imposed by
the LEP data as a function of the bilepton mass. The solid line 
corresponds to 2.7 fb$^{-1}$ of integrated luminosity  collected at
energies ranging from 130 GeV up to 206 GeV, by the four LEP
collaborations. We also show the exclusion areas in the $\lambda$--$M_B$
parameter space imposed by $M$--$\overline{M}$ conversion experiments. 
The limits from the LEP data on $\tilde{\lambda}_1$, and the region 
already excluded by $M$--$\overline{M}$ conversion, corresponding to
$\tilde{\lambda}_1/M_B < 0.20$ TeV$^{-1}$ are not shown in Fig.\ \ref{LEP}
but can be readily obtained from the limits for $\lambda_3$ using the fact
that $\tilde{\lambda_1} = \sqrt{2} \lambda_3$. In order to obtain these
results we have taken into account an efficiency of $90\%$ for dimuon 
reconstruction and a geometric acceptance of $|\cos \theta|  < 0.9$. This
was a conservative choice for the geometric acceptance since we assumed
the L3 value and all other experiments have larger acceptances. The
maximum allowed values of $\lambda/M_B$ at LEP are smaller than those
obtained from muonium--antimuonium conversion by a factor of two for most
of the bilepton mass range.

\section{Search at Future Linear Colliders}
 \label{sec:nlc}
\subsection{Electron-Positron Collider} 

Amid the efforts that are under way to determine the potentiality of
proposed linear colliders to discover new physics \cite{Peskin}, studies
on bilepton searches have been carried out mostly for its $e^- e^-$
collider operation mode \cite{Cuypers,Frampton,Raidal}.  We concentrate
here on the potential of the $e^+ e^-$ and $e \gamma$ operation modes to
discover doubly charged bileptons. We have assumed some representative
values for the energy and luminosity of new electron--positron machine
\cite{colliders}, namely $\sqrt{s} = 500$, $800$, and $1000$ GeV with
${\cal L} = 500$ fb$^{-1}$. This integrated luminosity is expected to be
achieved in one or two years of the collider operation.

We evaluated the $e^+ e^-$ operation mode in a similar way we did for LEP.
In Fig.\ \ref{cplotee}, we plot the $95 \%$ confidence level discovery
region in the $\lambda$--$M_B$ parameter space. In this figure, we also
plotted the region of the parameter space already excluded by
muonium--antimuonium conversion experiments. We considered a $90\%$ dimuon
reconstruction efficiency.  We did not consider a particular value for the
geometric acceptance, since we estimated that a $5^\circ$ aperture in the
beam pipe region would not lead to more than $2\%$ event loss.

\begin{figure}
\centering
\mbox{\epsfig{figure=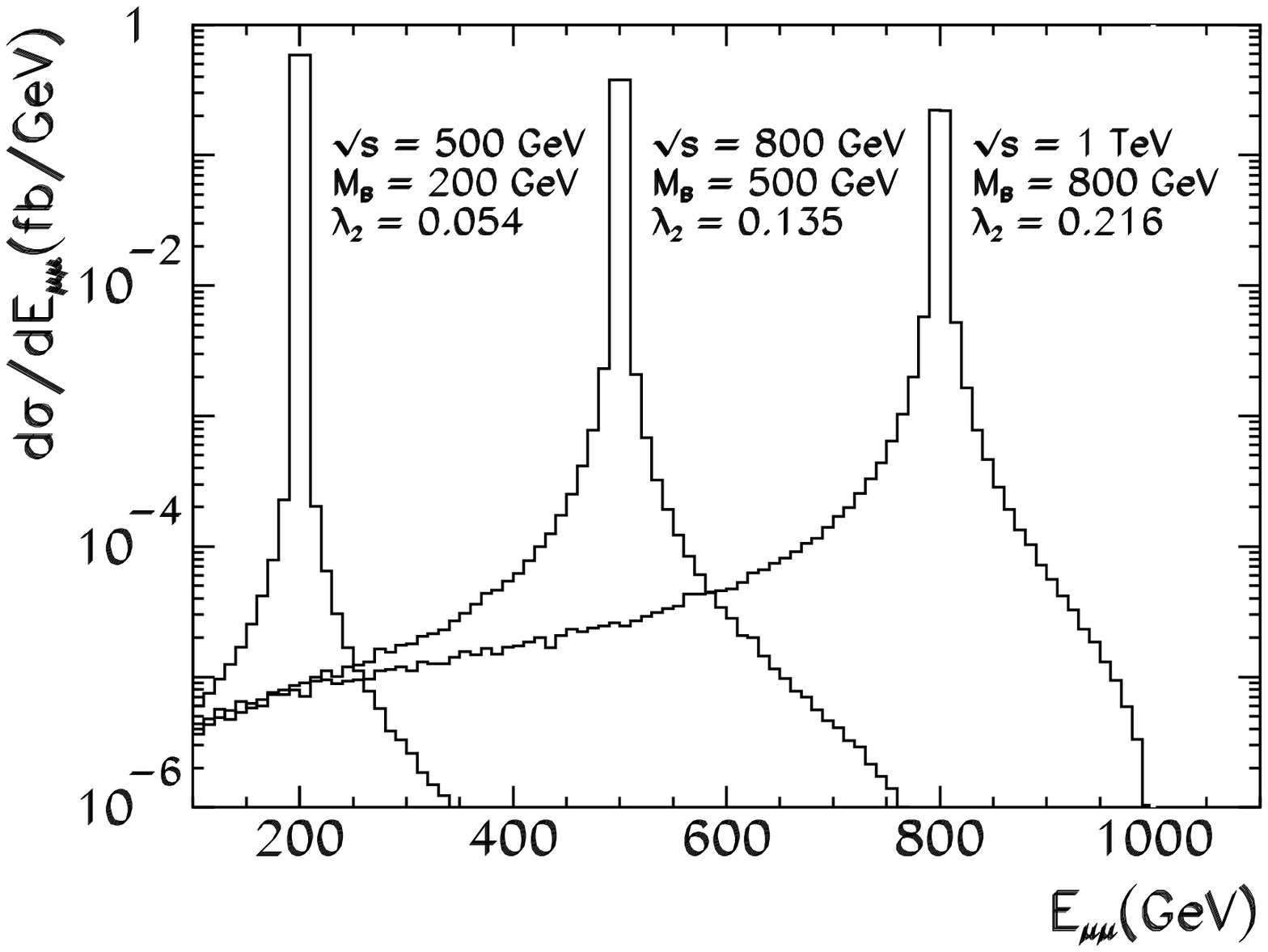,width=0.95\linewidth,%
      bbllx=20,bblly=210,bburx=530,bbury=590}}
\caption{Cross-sections for the $\mu^\pm \mu^\pm$ production in $e^+ e^-$
colliders as a function of their total invariant mass $E_{\mu \mu}$.} 
\label{eeres}
\end{figure}

If bileptons are indeed observed, and profusely produced, it will be
possible to determine their mass, and to learn whether they are vector or
scalar particles.  In Fig.\ \ref{eeres}, we present the expected dimuon
invariant mass differential cross section, for some specific values of the
vector bilepton coupling, which shows the resonant peak on the bilepton
mass. We have chosen for $\lambda_2$ the maximum value allowed by the
present $M$--$\overline{M}$ limits, {\it i.e.}, $\lambda_2 = 0.27\; M_B$.

The discrimination between vector and scalar can be done based on the fact
that the cross section for the process $e^- e^- \rightarrow \mu^- \mu^-$
changes with the polarization of the initial electron beams according to
which of the three  kind of bileptons participates in the reaction. The
discrimination can also be done using the fact that a vector bilepton
leads to an angular distribution of the final state muons that is
different from the angular distribution that would be observed if
bileptons were scalars. The dependences on the initial beam polarizations
($P_1$ and $P_2$) are shown in Tab.\ \ref{deptab}, as well as the
dependence on the angular distribution of the produced muons.

A discrimination based on the initial beam polarization is very promising
since in the next generation of linear colliders one expects up to $80 \%$
of polarization for the electron beam. Assuming that the electron emitted
from the positron beam is kept unpolarized ($P_2=0$), we can see that, as
the polarization of the electron beam ($P_1$) increases, the cross section
for the process  mediated by $\tilde{L}_1^{--}$ increases while the cross
section for the process mediated by $L_3^{--}$ decreases. On the other
hand, the insensitivity to $P_1$ would indicate the presence of a vector
bilepton, as can be seen from Tab.\ \ref{deptab}. However, if 
$\tilde{L}_1^{--}$ and $L_3^{--}$ get mixed, the dependence on $P_1$ can
be too small ($\tilde{\lambda_1} = \sqrt{2}\lambda_3$) to be observed, and
it would not be possible to distinguish between vectors and scalars by
looking at the electron beam polarization dependence of the cross-section.

\begin{figure} 
\centering 
\mbox{\epsfig{figure=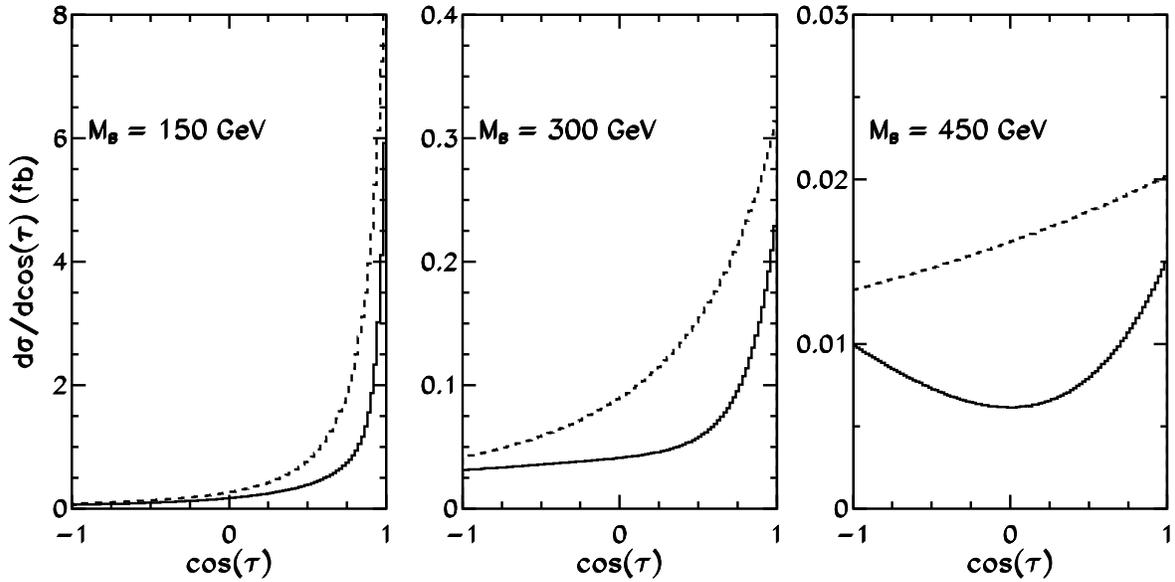,width=0.95\linewidth,%
      bbllx=45,bblly=280,bburx=530,bbury=530}} 
\caption{Angular distribution in laboratory frame at $\sqrt{s} = 500$ GeV. The
solid (dashed) line corresponds to vector (scalar) bileptons.}  
\label{ang_dependence} 
\end{figure}

The alternative would be to examine the angular distribution of the final
state muons. In Fig.\ \ref{ang_dependence} we can see the difference
between the angular distribution of reactions mediated by scalar and
vectorial bileptons at  $\sqrt{s} = 500$ GeV.  However, when the bilepton
mass is relatively small, like for a $150$ GeV bilepton depicted in Fig.\
\ref{ang_dependence}, the angular distributions for vector and scalar
bileptons are quite similar. In this case, to distinguish vector from
scalar particle it will require the knowledge of the bilepton mass with
good accuracy. Fortunately, as can be seen from Fig.\
\ref{ang_dependence}, the smaller the bilepton mass the bigger is the
production cross-section. This may help to make the identification task
easier since more events are expected in this case for a given value of
the coupling constant.

\subsection{Electron-Photon Collider}
 
Projects for future linear colliders include the possibility to transform
the initial $e^+ e^-$ machine into an $e \gamma$ or $\gamma \gamma$
collider with comparable energy and luminosity \cite{egammanew}. In these
machines, a highly energetic  photon beam is generated by Compton
backscattering  low energy photons emitted by a laser. The high energy
backscattered photons are then made to collide with the opposite incoming
beam. For a non--polarized collider, the backscattered photon distribution
function is given by \cite{egammaold}, 
\begin{equation}
f^\gamma_e(x)=\frac{2}{\sigma^0}\left[\frac{1}{1-x}+1-x-4r(1-r)\right] ,
\label{spectrum}
\end{equation}
where $x$ is the fraction of the electron momentum carried by the photon,
$r = x/[y(1-x)]$, and
\begin{equation}
\sigma^0 = \left( 2 - \frac{8}{y} - \frac{16}{y^2} \right) \ln(y + 1)
+ 1 +  \frac{16}{y} - \frac{1}{(y + 1)^2} \;,
\end{equation}
where $y \approx 15.3\,E_B\,\omega_0$, with the parent electron energy
$E_B$ expressed in TeV, and the laser energy $\omega_0$ in eV. A maximum
value $y=4.8$ is usually adopted to avoid electron regeneration  through
pair production.  This is the $y$ value we used in our analysis.

\begin{figure}
\centering
\mbox{\epsfig{figure=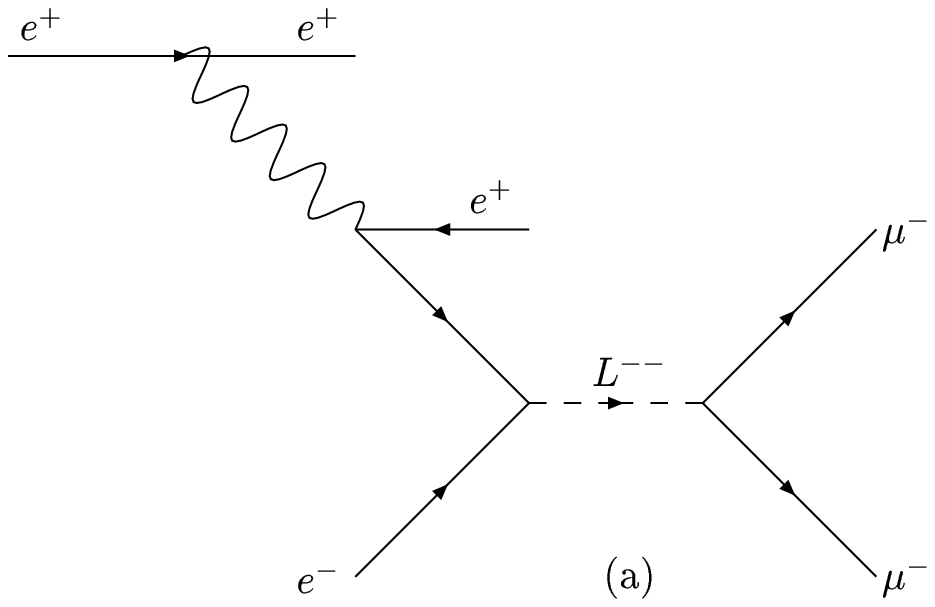,width=0.48\linewidth,%
      bbllx=120,bblly=500,bburx=400,bbury=680}}
\hfill 
\mbox{\epsfig{figure=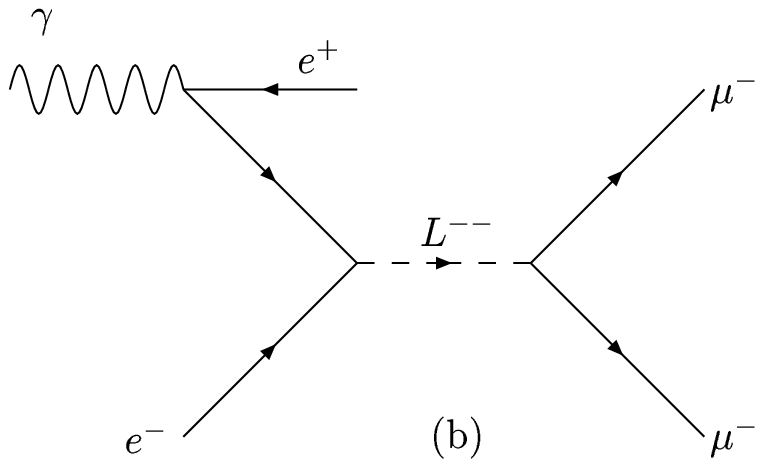,width=0.48\linewidth,%
      bbllx=120,bblly=500,bburx=400,bbury=680}} 
\caption{Main contribution to the processes (a) $e^+  e^- \rightarrow
\mu^{\pm} \mu^{\pm}$ and (b) $e^- \gamma \rightarrow
\mu^- \mu^-$.}
\label{eefig}
\end{figure}

In the $e \gamma$ collider, bileptons would be produced as a $s$-channel
resonance through the diagram depicted in Fig.\ \ref{eefig}(b). Similarly
to the treatment we employed for $e^+ e^-$ colliders, we also assumed here
that the positron escapes unobserved down the beam pipe.  The
cross-section for the process $e^- \gamma \rightarrow \mu^- \mu^-$ in the
equivalent particle approximation is given by,
\begin{equation} 
\sigma_{e^- \gamma \rightarrow \mu^- \mu^-} =
\int_{x_1^{\rm min}}^{x_1^{\rm max}} \int_{x_2^{\rm min}}^1 dx_1 \, 
dx_2 \,f^{\gamma}_e(x_1)\, F_{\gamma}^e(x_2)\,
\hat\sigma_{e^- e^- \rightarrow \mu^- \mu^-}
\end{equation}
where $f^{\gamma}_e(x_1)$ is given by Eq.\ (\ref{spectrum}), and 
$F_{\gamma}^e(x_2)$ is the distribution function of the equivalent 
electron carrying a fraction $x_2$ of the photon energy, 
\begin{equation}
F_{\gamma}^e(x_2) = \frac{\alpha}{\pi}\left[ x_2^2 + (1-x_2)^2 \right]
\ln\frac{E_\gamma}{m_e} \; .
\label{egammadist}
\end{equation}

We also determined the potential of the  $e \gamma$ collider to discover
bileptons considering the observation of a single event. It has been
assumed that the luminosity of the $e \gamma$ mode will be comparable to
its parent $e^+ e^-$ mode.  In Fig.\ \ref{cploteg} we plot the $95 \%$
confidence level discovery region in the $\lambda$--$M_B$ parameter space.
In this figure, we also plotted the region of the parameter space already
excluded by muonium--antimuonium conversion experiments. We can see that
an $e \gamma$ collider is more efficient for the search of bileptons than
the $e^+ e^-$ collider.  This is expected since the laser backscattering
mechanism (Fig.\ \ref{eefig}(b)) is capable to generate a large amount of
hard photons when compared with the usual bremsstrahlung subprocess (Fig.\
\ref{eefig}(a)).

\begin{figure}
\centering
\mbox{\epsfig{figure=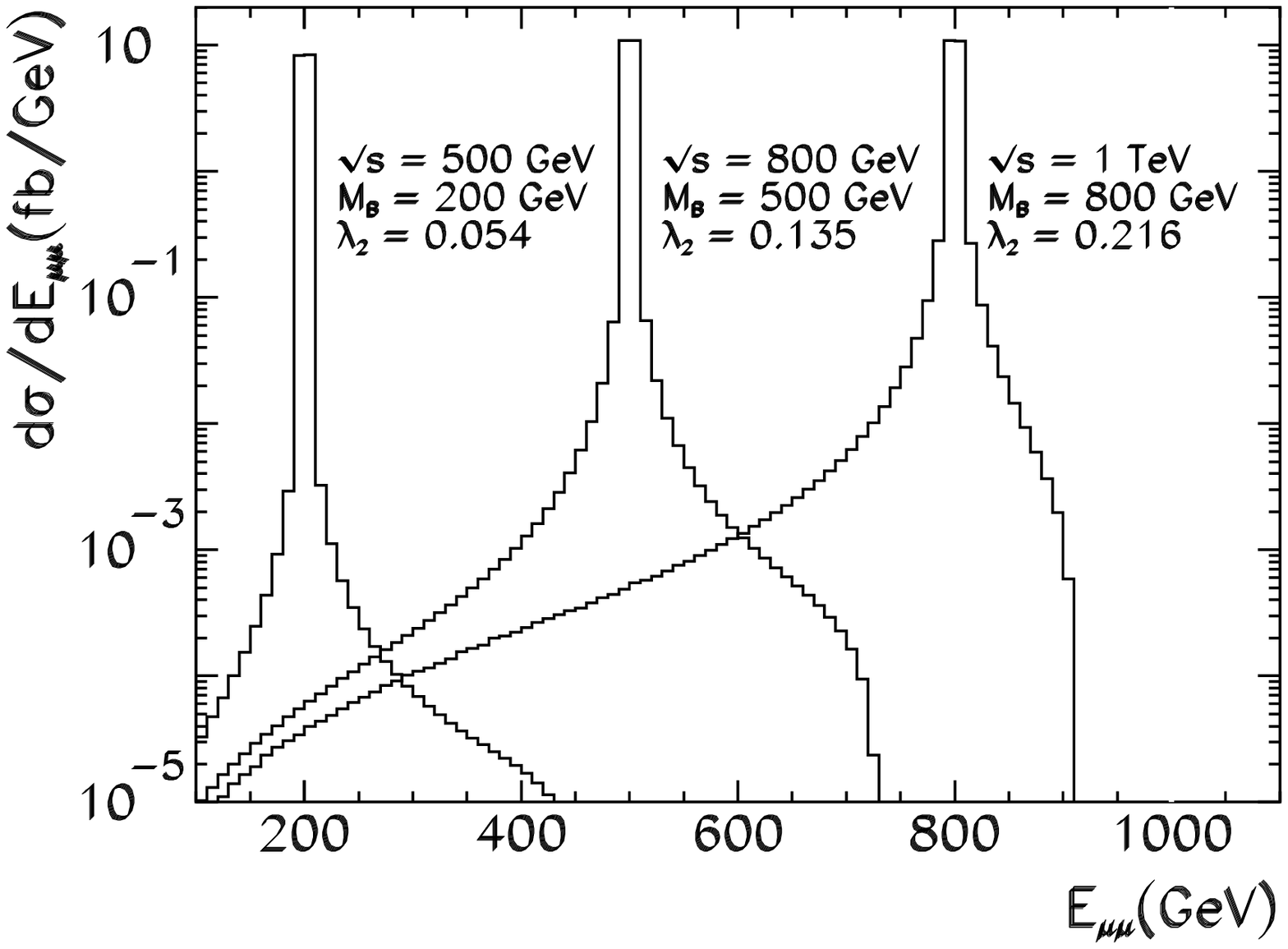,width=0.95\linewidth,%
      bbllx=25,bblly=210,bburx=525,bbury=580}}
\caption{Cross sections for the $\mu^\pm \mu^\pm$ production in $e
\gamma$ colliders as a function of their total invariant mass $E_{\mu \mu}$. } 
\label{egres}
\end{figure}

If the bileptons are discovered and a fairly large amount of events is
collected, we can use the same procedure, based on the electron
polarization and angular distribution of final state muons, to determine
their mass and spin. In Fig.\ \ref{egres}, we present the expected event
distribution as a function of the muon (or antimuon) pair invariant
mass.   As in Fig.\ \ref{eeres} the value of $\lambda$ is the maximum
allowed by the present $M$--$\overline{M}$ limits, {\it i.e.}, $\lambda_2
= 0.27\; M_B$. The angular distribution follows the same pattern as the
one presented in Fig.\ \ref{ang_dependence}.

 
\section{Discussions and Conclusions}
\label{conclusions}

\begin{figure}
\centering
\mbox{\epsfig{figure=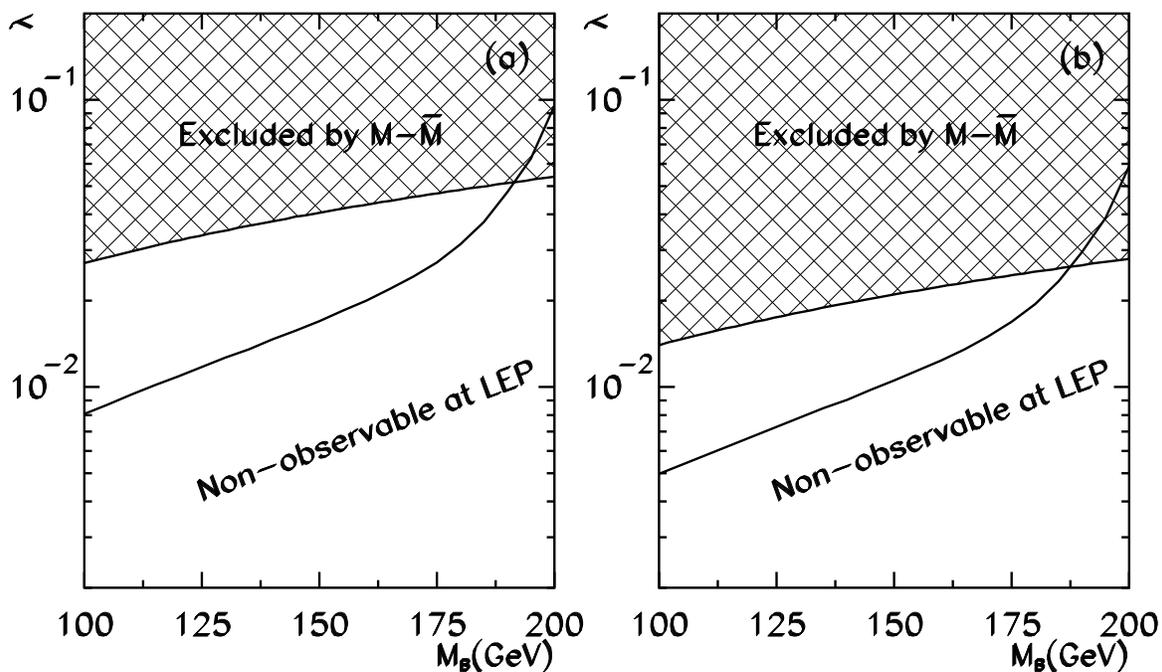,width=0.95\linewidth,%
      bbllx=50,bblly=260,bburx=550,bbury=550}}
\caption{Bilepton exclusion plot in the $(M_B, \lambda)$ plane for
LEP data. (a) limits on $\lambda_2$ (95\% C.L.). (b) limits on $\lambda_3$ (90\%
C.L.). See comments on Sec.\ \ref{sec:lep} for the limits on
$\tilde{\lambda}_1$.} 
\label{LEP}
\end{figure}

\begin{figure} 
\centering 
\mbox{\epsfig{figure=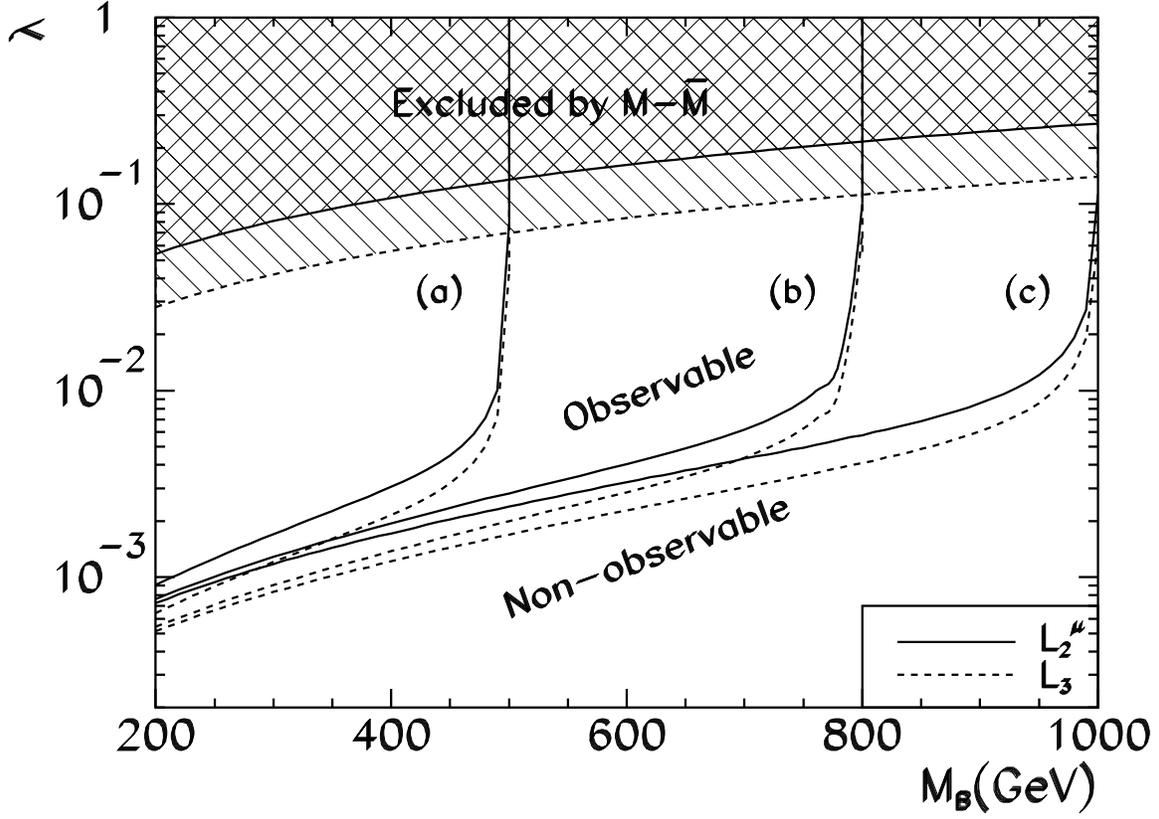,width=0.95\linewidth,%
      bbllx=30,bblly=210,bburx=550,bbury=590}}
\caption{Discovery region in the $(M_B, \; \lambda)$ plane at $95 \%$
C.L., assuming ${\cal L}$ = 500 $fb^{-1}$ and $\sqrt{s} = 500$ (a), 800
(b), and 1000 (c) GeV, for a future $e^+ e^-$ linear collider.}  
\label{cplotee}
\end{figure}

\begin{figure}
\centering
\mbox{\epsfig{figure=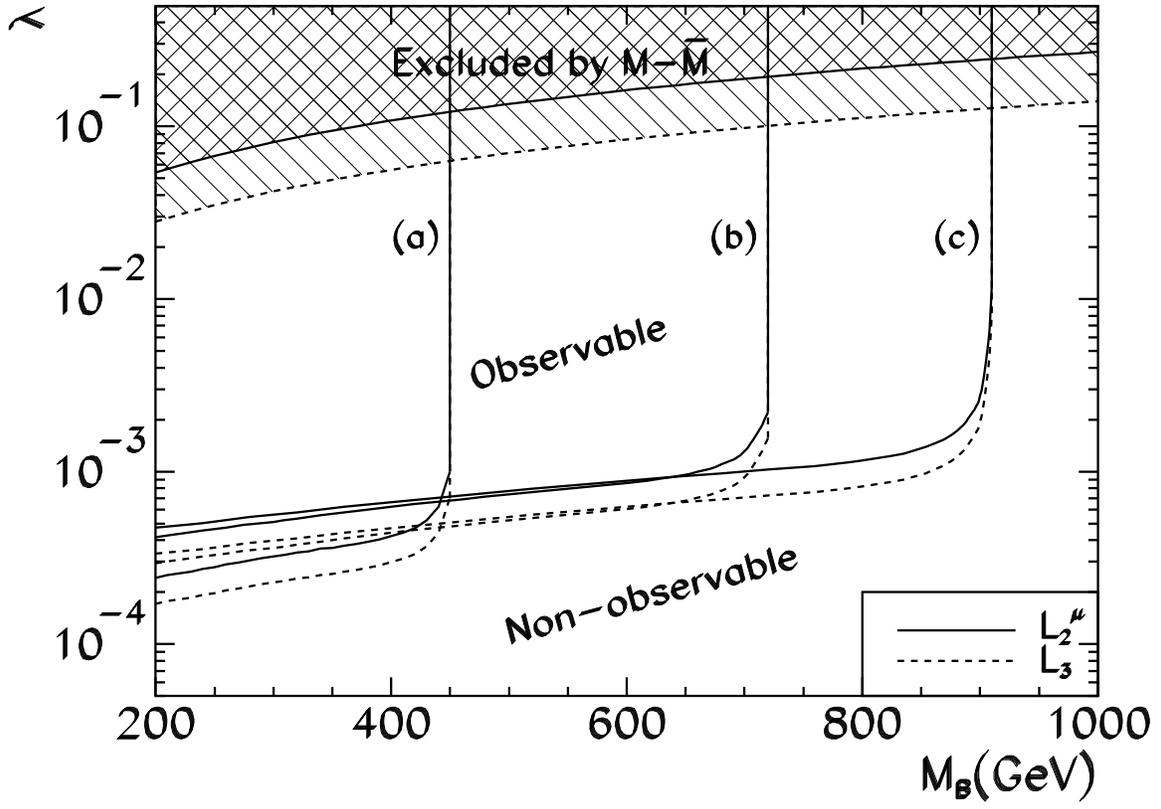,width=0.95\linewidth,%
      bbllx=30,bblly=210,bburx=550,bbury=580}}
\caption{The same as Fig.\ \ref{cplotee} for the $e \gamma$ mode of the
linear collider.}  
\label{cploteg}
\end{figure}

In this article we have shown that new and more stringent limits on the
coupling constants and masses of doubly charged bileptons can be obtained
from LEP-II data. As for the limits from $M$--$\overline{M}$ conversion,
our results apply to the case of flavor--diagonal bilepton couplings.
These stringent bounds result from the fact that the integrated luminosity
collected by the LEP Collaborations at energies above 189 GeV is very
large. In order to obtain the new limits on $\lambda$ and $M_B$ we have
applied the equivalent particle approximation method. It is expected that
an exact tree level calculation of the cross-section would lead to only
minor corrections to our results \cite{EPA}. A few percent uncertainty in
the cross-sections translates into even smaller corrections to the curves
in Figs.\ \ref{LEP}, \ref{cplotee} and \ref{cploteg} because the
cross-sections are proportional to $\lambda^2$. Our results suggest that
an accurate analysis taking into account detailed LEP detectors and data
sample properties will be able to establish bounds on bilepton couplings
and masses that overcome the present ones.

We have also shown that the construction of high energy linear
accelerators could lead to a large increase in the sensitivity to the
doubly charged bileptons. If these new particles are in the reach of these
machines, a quite evident resonance peak should be observed without the
need to scan the collision energy. The sensitivity to bileptons of the
general purpose $e \gamma$  collider is of the same order of magnitude as
the $e^- e^-$ operating mode. In order to compare both operation modes we
updated the results presented at Ref. \cite{Cuypers2} on bilepton search
in $e^- e^-$ colliders, to the luminosity of 500 fb$^{-1}$ considered
here. The reanalysis is straightforward and is based on the fact that the
cross-section for the process $e^- e^- \rightarrow \mu^- \mu^-$ is
proportional to the square of the bilepton coupling and the limits on
$\lambda$ will be proportional to $\sqrt{1/\cal L}$, where $\cal L$ is the
collider luminosity. In Tab.\ \ref{compare} we compare the limits on
$\lambda_2$ for the two kind of colliders operating at $\sqrt{s} = 0.5$
and 1 TeV,  with ${\cal L} = 500$ fb$^{-1 }$.  As it would be expected,
the $e^- e^-$ mode is, in general, more sensible than the $e \gamma$ mode.
This advantage may be however illusory. The instantaneous luminosity
delivered by the $e^- e^-$ collider is expected to be half that of the
$e^+ e^-$ with same beam characteristics because of the anti-pinch effect.
This fact slightly diminishes the advantage of the $e^- e^-$ mode over the
$e \gamma$ mode for bilepton production. By comparing Fig.\ \ref{cplotee}
with the values in Tab.\ \ref{compare}, we can see that the direct $e^+
e^-$ mode is clearly less sensitive than the other two operation modes. 
Nevertheless, it is  much more sensitive than the present experiments.

In conclusion, a collider operating at $e^+ e^-$ or $e \gamma$ modes is
quite sensitive to the resonant production of doubly charged bileptons. If
bileptons are indeed  observed in these machines we will probably be able
to get a lot of information about then. Later, their properties could  be
thoroughly  studied using the $e^- e^-$ mode operating at $\sqrt{s} =
M_B$.


\acknowledgments

E.M.G.\ is grateful to University of Wisconsin for its kind hospitality. 
A.G. would like to thanks the hospitality at the Instituto de Fisica
Corpuscular (IFIC-UV) were part of this work was carried on. S.F.N.\ is
grateful to Fermilab for its hospitality. This work was supported by
Conselho Nacional de Desenvolvimento Cient\'{\i}fico e Tecnol\'ogico
(CNPq), by Funda\c{c}\~ao de Amparo \`a Pesquisa do Estado de S\~ao Paulo
(FAPESP), by Programa de Apoio a N\'ucleos de Excel\^encia (PRONEX), and
by Funda\c{c}\~ao para o Desenvolvimento da UNESP (FUNDUNESP). A.G. was
partially supported by Funda\c{c}\~ao Coordena\c{c}\~ao de Aperfei\c{c}oamento
de Pessoal de N\'{\i}vel Superior (CAPES).



\begin{table}
\begin{tabular}{lcccccccccc}
$\sqrt{s}$ (GeV)  & 133 & 161 & 172  & 183 & 189 & 192 &
196 & 200 & 202 & 206 \\
${\mathcal{L}}(\sqrt{s})$ (pb$^{-1}$) & 22 & 42 & 41 &
217 & 678 & 113 & 313 & 328 & 155 & 800 
\end{tabular}
\caption{LEP integrated luminosities at different energies. The luminosity at  $\sqrt{s} = 133$ GeV is the  weighted average of the luminosities
obtained at 130 GeV and 136 GeV.}
\label{tableLEP}
\end{table}

\begin{table} 
\begin{tabular}{lcc} 
 & Polarization dependence & Angular dependence \\ \hline   
 $\tilde{L}_1^{--}$ & $(1 + P_1)(1 + P_2)$ & none \\
$L_{2\mu}^{--}$ & $1 - P_1 P_2$ & $1 + \cos^2
\theta^\ast$ \\  
$L_3^{--}$ & $(1 - P_1)(1 - P_2)$ & none \\ 
\end{tabular} 
\caption{Angular and polarization dependence for $e^+ e^- \rightarrow
\mu^- \mu^-$ ($\theta^\ast$ is the C.M.\ angle).} 
\label{deptab}
\end{table} 

\begin{table} 
\begin{tabular}{||c|c|c||c|c||} 
$M_B$ & \multicolumn{2}{c||}{$\lambda_2 \times 10^4$ (500 GeV)} & 
        \multicolumn{2}{c||}{$\lambda_2 \times 10^4$ (1 TeV)} \\
	\hline
      & $e^- e^-$  &  $e \gamma$ & $e^- e^-$  &   $e \gamma$ \\
      \hline
 200  & $2.50$ &  $2.40$  &  $1.46$ &  $4.72$ \\
 300  & $0.96$ &  $3.22$  &  $1.36$ &  $5.55$ \\
 400  & $0.60$ &  $4.21$  &  $1.23$ &  $6.64$ \\
 500  & $0.11$ &  ---     &  $1.10$ &  $7.73$ \\
\end{tabular} 
\caption{Values of $\lambda_2$ that lead to the expected number of three
events in the case ${\cal L} = 500$ fb$^{-1}$.} 
\label{compare}
\end{table} 


\end{document}